# New phase transition in $Pr_{1-x}Ca_xMnO_3$ system: evidence for electrical polarization in charge ordered manganites

A.M.L. Lopes[1, 2, 3 *], J.P. Araújo[2], V.S. Amaral[3], J.G. Correia[1,4], Y.Tomioka[5] and Y. Tokura[6]

[1] CERN EP, CH 1211 Geneva 23, Switzerland
[2] Departamento de Física, IFIMUP, Universidade do Porto, 4169-007 Porto, Portugal.
[3] Departamento de Física and CICECO, Universidade de Aveiro, 3810-193 Aveiro, Portugal.
[4] Instituto Tecnólogico Nuclear, E.N. 10, 2686-953 Sacavém, Portugal.
[5] CERC, National Institute of Advanced Industrial Science and Technology, Tsukuba, Ibaraki 305-8562, Japan
[6] Department of Applied Physics, University of Tokyo, Tokyo, 113-8656, Japan
*: corresponding author (armandina.lima.lopes@cern.ch)

A detailed study of the electric field gradient (EFG) across the $Pr_{1-x}Ca_xMnO_3$ phase diagram and its temperature dependence is given. Clearly, distinct EFG behaviours for samples outside/inside the charge order (CO) region are observed. The EFG temperature dependence evidences a new phase transition occurring over the broad CO region of the phase diagram. This transition is discontinuous and occurs at temperatures between the charge ordering and the Néel temperatures. The features observed in the EFG are associated with polar atomic vibrations which eventually lead to a spontaneous local electric polarization below CO transition.

The exquisite coupling between lattice, spin, charge and orbital degrees of freedom, that led to renowned phenomena like high-$T_c$ superconductivity, colossal magnetoresistance and multiferroic behaviour, still challenges our understanding of transition metal oxides [1]. In $Mn^{3+}/Mn^{4+}$ mixed valence manganites this subtle entanglement of the several degrees of freedom brings about competing orbital, magnetic, and dielectric orders depending on the doping, temperature, and external stimulation. In particular, much attention has been devoted to the charge and orbital ordered (CO, OO) phases, *i.e.*, a real-space ordering of charge and orbitals due to the electron-phonon and long-range Coulomb interactions. The classic CO picture with a $Mn^{3+}$- $Mn^{4+}$ *checkerboard pattern* [2] has been questioned [3,4] since the work of Daoud-Aladine *et al*. [5]. These authors proposed an electronic ground-state where one $e_g$ electron is shared by two $Mn^{3+}$ ions, the so-called *bond-centred Zener polaron picture*. Subsequently, Efremov *et al*. [6] proposed a new scenario where the *bond-centred* ($Mn^{3+}$-$O^-$-$Mn^{3+}$ dimmers) and the *site-centred* CO pictures coexist and the result breaks the inversion symmetry, leading to the appearance of a spontaneous electric polarization. More recently, it has been demonstrated that a commensurate spin-density-wave ordering with a phase dislocation can also give rise to a polar ferroelectric distortion in rare-earth manganites [7]. In a different context, a frustrated CO state was also shown to lead to an electrical polarization in $LuFe_2O_4$ [8]. Although the CO state in $Pr_{1-x}Ca_xMnO_3$ is currently referred to as a new paradigm for ferroelectrics [9, 10, 11], it has been very hard to prove that electric polarization exists in CO $Pr_{1-x}Ca_xMnO_3$ and in similar CO manganites [9, 10, 12]. This is connected to the relatively high conductivity of these materials, and to the possibility that the suspected electric dipole order may only occur within nanoscopic regions. However, a very recent work of Ch. Jooss *et al* [13] provides, by refinements of electron diffraction microscopy data, indirect evidence for canted antiferroelectricity in $Pr_{0.68}Ca_{0.32}MnO_3$.

The measurement of the electric field gradient tensor (EFG) via hyperfine techniques offers a very sensitive tool to locally study phase transitions and probe electric ordering [14, 15]. Aiming to get further insight on the microscopic nature of CO we performed a detailed study on the electric field gradient temperature trends across the $Pr_{1-x}Ca_xMnO_3$ phase diagram. This study reveals clearly distinct EFG behaviours for samples outside/inside the CO region. Through our measurements evidence for a phase transition occurring between the charge ordering and the Néel temperatures are forwarded. This transition appears in samples over the broad CO phase diagram region and as the composition shifts away from x=0.5, the new critical temperature lowers away from the CO temperature. A local paraelectric susceptibility was evidenced through the EFG principal component, $V_{ZZ}$, and shows a sharp increase towards the observed transition. In this way, these experimental results hint for the predicted paraelectric to ferroelectric phase transition in the CO $Pr_{1-x}Ca_xMnO_3$ manganites.

$Pr_{1-x}Ca_xMnO_3$ polycrystalline samples were produced by the solid state reaction method. The sample's crystallographic structure and lattice parameters were determined by refinement of x-ray powder diffraction patterns. The temperature dependence of magnetization was used to determine the characteristic Néel($T_N$), Curie ($T_C$), and charge-order ($T_{CO}$) temperatures. For all compositions the system presents an orthorhombic distorted perovskite structure and CO is stabilized over a broad range of compositions, with x extending from 0.30 to 0.90, in good agreement with literature [16, 17, 18]. Using the $^{111m}Cd$ γ-γ perturbed angular correlation technique (PAC) [19] the measurement of the EFG tensor at the Ca/Pr site was performed in a series of samples ranging from x=0 to 1, in the 10 K to 1000 K temperature range (for typical experimental details see [15]). To perform PAC measurements the samples under study were implanted at ISOLD/CERN with $^{111m}Cd$. Note that the density of probing Cd atoms is lower than 1 ppm of the Pr/Ca concentration. The fit to the PAC experimental anisotropy function, R(t), was performed by the numerical diagonalization of the interaction Hamiltonian for a static electric quadruple interaction, which, in the proper reference frame of the EFG tensor, with $|V_{ZZ}|\geq|V_{YY}|\geq|V_{XX}|$, reads:

$$H = \frac{eQV_{zz}}{4I(2I-1)\hbar}\left[3I_Z^2 - I(I+1) + \frac{1}{2}\eta(I_+^2 + I_-^2)\right]$$

where I=5/2 is the nuclear spin of the probe and $\eta = (V_{XX} - V_{YY})/V_{ZZ}$ is the EFG axial symmetry parameter. The anisotropy function may be expanded as $R(t) = \sum A_{kk} G_{kk}(t)$ with $A_{kk}$ being the angular correlation coefficients of the nuclear decay cascade and $G_{kk}(t)$ contains the signature of the lattice fields interacting with the probes. In its diagonal form the EFG tensor is fully characterized by the principal component, $V_{ZZ}$, and axial symmetry η parameters [19]. For all studied compositions, the fit to each PAC spectrum was performed considering only one main Lorentzian-like EFG distribution (see Fig.1 for representative PAC spectra).

The study of the temperature dependence of the EFG parameters, over the whole composition range, revealed a slightly decreasing asymmetry parameter without any unusual features (insets of Fig. 2), whereas the principal component of the EFG, $V_{ZZ}$, revealed a rich variety of behaviours. Samples outside the CO region of the phase diagram showed an increase in $V_{ZZ}$ with decreasing temperature without any noticeable anomaly (see Fig. 2a for x=0.25). In perovskite related systems the EFG temperature dependence is essentially due to EFG fluctuations due to lattice vibrations. This issue was studied in the past and the EFG temperature dependence emerged as a probe of lattice vibrations, particularly useful to signalise (anti)ferroelectric transitions [20,21]. In fact, expanding the EFG, in particular its principal component, $V_{ZZ}$, in powers of the atomic displacements, two main contributions emerge after performing the time average: the first arises from the rigid lattice, $V_{ZZ}^0$, while the second comes from the atomic mean square displacements [21], $\langle\zeta^2\rangle_t$, i.e. $\langle V_{ZZ}\rangle_t \sim V_{ZZ}^0 + \beta\langle\zeta^2\rangle_t$. Accordingly, $V_{zz}$ follows closely the temperature dependence of the mean square displacements. For samples outside the CO region the $V_{zz}$ temperature evolution was modelled considering the Planck oscillator with an average frequency for the rotation modes around the $V_{zz}$ axis (the stretching vibrations contribution, with higher frequencies, is negligible with respect to rotation modes) [21]. The fit of the experimental data held an energy of 20(1) meV (158(9) cm$^{-1}$) and a high temperature asymptotic linear slope of -1.5(1)×10$^{-4}$ K$^{-1}$ (normalised to the value of the rigid lattice neglecting the zero point fluctuations). In contrast to the previous case, samples within the CO region present a very unusual $V_{ZZ}$ thermal dependence (see Fig. 2b for x=0.85). For these samples, the common negative linear slope of the $V_{ZZ}(T)$ trend is only observed much above 300 K. Below that temperature one sees that $V_{ZZ}$ decreases with decreasing temperature until the charge-order temperature is reached.

The distinct $V_{ZZ}$ behaviour at room temperature, for samples within and outside the CO region, is patent in Fig. 3 where the $V_{ZZ}$ dependence on the Ca content is shown. Increasing x, near x=0.30, the observed changes in $V_{ZZ}$ are connected with the change in the $V_{ZZ}(T)$ slope from negative values to positive ones and suggest distinct local dynamics for compositions within the CO region of the phase diagram. Note that at 896 K (inset of Fig. 3) only a linear $V_{ZZ}$ decrease with Ca content increase is observed. The mentioned anomalous decrease in $V_{ZZ}$ with decreasing temperature, can be easily understood if in addition to the harmonic lattice vibrations model, considered above, one adds a softening in a rotational mode (with a frequency softening $\omega_0^2 \sim (T-T_c)$) [20]. The fit to the data above CO temperature held a asymptotic linear slope of -1.5(2)×10$^{-4}$ K$^{-1}$, i.e., similar to the one obtained for samples outside the CO region, and allowed to roughly estimate the transition temperature $T_C$~239 K ($T_{CO}$=235 K) for x=0.35, $T_C$~236 K ($T_{CO}$=240 K) for x=0.40, $T_C$~278 K ($T_{CO}$=274 K) for x=0.65 and $T_C$~147 K ($T_{CO}$=151 K) for x=0.85, i.e., close to the CO temperature (deduced from the magnetic susceptibility measurements shown in Fig. 4). A similar anomalous decrease in $V_{ZZ}$ was also found in high-T*c* superconductors and attributed to anisotropic lattice displacements connected to the softening of oxygen vibrational modes [22]. Thus, our results suggest that anisotropic lattice vibrations are a precursor effect of the charge ordering. Our data thus strongly support that the softening of vibrational modes towards the CO transition results in a lattice instability near $T_{CO,}$ as suggested by recent ultrasounds studies in $Pr_{1-x}Ca_xMnO_3$ compounds, that found evidence for several types of elastic softening above the CO phase transition [23].

A key feature of this work is the finding of a sharp rise of $V_{ZZ}$ followed by a discontinuity (see fig. 2b), observed when lowering temperature below $T_{CO}$. Such prominent EFG variation indicates the existence of a phase transition and clearly the atomic displacement thermal fluctuations enhance below the $T_{CO}$, preceding a discontinuous phase transition. The question that now arises is, what is the origin of this phase transition?

According to Milward *et al* [24] for x<0.5 and slightly hole doped CO systems should present a incommensurate(IC)-commensurate(C) phase transition at a temperature below CO while highly hole doped CO manganites are expected to be incommensurate down to the lowest temperatures. In fact, from diffraction data, a lock-in transition was found near $T_N$ for x=0.5, both for the Pr-Ca and the La-Ca systems [25, 26]. Moreover, very recently, in certain strain conditions, a commensurate state in $Pr_{0.48}Ca_{0.52}MnO3$ was observed [27]. CO $Pr_{1-x}Ca_xMnO_3$ compounds are generally considered to be commensurate for x<0.5. Is it possible that the transition observed in our data is a lock-in transition as discussed in literature? Such IC-C transition if present should be revealed by different EFG distribution profiles above and below the transition (and also at the onset of CO). Remark that in an IC phase the number of non-equivalent lattice sites is infinite and therefore the resulting EFG distribution evidence strongly non-lorentzian character, usually being broad and asymmetric with respect to the distribution peaks. On the contrary, our PAC signal always show the typical I=5/2 frequency triplets with the expected amplitude ratios (and broadening ratios) that can be properly fitted with an Lorentzian-like distribution in the whole studied temperature range. In our view, this alone strongly suggests that our signal arises from a commensurate (or normal) phase. We cannot exclude, however, the presence of incommensurate (narrow) solitonic phases as predicted by Brey and Littlewood [28]. In this scenario probes at regions outside the soliton would contribute for the PAC modulation whereas probes within the soliton, thus within a very broad EFG distribution, would contribute only to a reduction of the PAC anisotropy. In any case our EFG discontinuity arises from a commensurate (or normal) region and thus the origin of the observed phase transition has to be founded in other grounds than a IC-C phase transition.

In this system the orbital ordering appears concomitantly with the CO [29] and although non-polar fluctuations could arise from orbital fluctuations of the $Mn^{3+}$ d-electrons they cannot also justify the experimental observations at T< $T_{CO}$.

On the other hand, fluctuations in the electric dipole moment induce similar features in the EFG [21] as the ones observed here and leaving unchanged the EGF distribution profile. In (anti)ferroelectric materials $\langle V_{ZZ}\rangle_t$ is commonly described by including a contribution from the static polar displacements of the ions/electrons, being thus connected with a local electric polarization P, and another from thermal fluctuations, being connected with the local electrical susceptibility $\chi$. Again, by expanding the $V_{ZZ}$ in powers of the atomic displacements and performing the time average, one finds: $\langle V_{ZZ}\rangle_t \approx V_{ZZ}^0 + \alpha P^2 + \beta T\chi$ [20, 21]. Here $\alpha$ and $\beta$ are scaling factors and $V_{ZZ}^0$ is now taken as $V_{ZZ}$ in the paraelectric phase considering no electric dipole fluctuations. Accordingly, one can attribute the sharp rise in $V_{ZZ}$ to polar fluctuations since $\chi \propto 1/(T-T_{EDO})$, where $T_{EDO}$ is defined as the critical temperature of the electric dipole order (EDO).

To model the $V_{ZZ}$ thermal dependence the Landau theory of phase transitions with a free energy power expansion up to sixth order in the polarization was used. The fits of $V_{ZZ}$ data are presented in Fig. 4 (d, e, f), together with the reciprocal magnetic susceptibility for three samples (d, e, f). As one can observe, the present phenomenological model matches the experimental data quite well, showing that these results are compatible with a scenario where electric dipoles appear below $T_{CO}$. The fit to the data below CO temperature held: $T_{EDO}$=206 K ($T_{CO}$=235 K) for x=0.35, $T_{EDO}$=218 K ($T_{CO}$ =240 K) for x=0.40, and $T_{EDO}$=112 K ($T_{CO}$ =151 K) for x=0.85. Clearly, one notices from these results that the phase transition occurs below $T_{CO}$. The obtained critical temperatures suggest that as the sample compositions drift away from x=0.5, the new critical temperature ($T_{EDO}$) drifts away from the CO temperature (see zone between vertical dashed lines in figures 4d, 4e, and 4f).

From our data it is not possible to obtain unambiguously a value for the spontaneous polarization neither to ascribe a ferroelectric or antiferroelectric nature of the observed phase transition, as the EFG can be also sensitive to the polarization of an electric dipole sublattice. Note that our data provides typical signatures of a phase transition and, consequently, the observed phenomenon cannot be a localised occurrence around the probe atom. Thermodynamics requires a collective state involving long-range ordering of local dipoles, though it can be on the order of only a few nm, according to recent results[30].Along these lines, our work has to be taken as an important ground of the growing experimental evidence of the existence of electric dipole order in CO manganites [11], [13].

Summarising, our results give experimental evidence for a new phase transition occurring below charge order transition in $Pr_{1-x}Ca_xMnO_3$ and one can draw conclusions on its first-order nature. The observed transition was interpreted in terms of a paraelectric to (anti)ferroelectric phase transition. Within our analysis, electric dipole order is not limited to the window below x=0.5 and spreads over the entire CO/OO region of the $Pr_{1-x}Ca_xMnO_3$ phase diagram.

This study thus raises the debate around the CO/OO nature in manganites, establishing new directions and setting boundaries where the CO/OO phases should be further investigated by the design of new experiments as well as by reassessing already existing data. It also raises theoretical challenges envisaging, in particular, the comprehension of the mechanisms that trigger the shift of this new transition from the CO temperature.

**Acknowledgements:** The authors gratefully thank P. B. Tavares, T. M. Mendonça, E. Rita, M. S. Reis, P. Pereira, T. Butz and the ISOLDE Collaboration for the preparation of some of the used samples and technical support. This work was funded by the EU (European Union Sixth Framework through RII3-EURONS, contract no. 506065 and Large Scale Facility contract HPRI-CT-1999-00018) and the Portuguese Research Foundation (FCT) and FEDER (projects FEDER/POCTI/n2-155/94, POCI/FP/63911/2005, POCI/FP/63953/2005, PDCT-FP-FNU-50145-2003, projects POCI/FP/63911/2005 and through the Associated Laboratory – IN).

# Figures

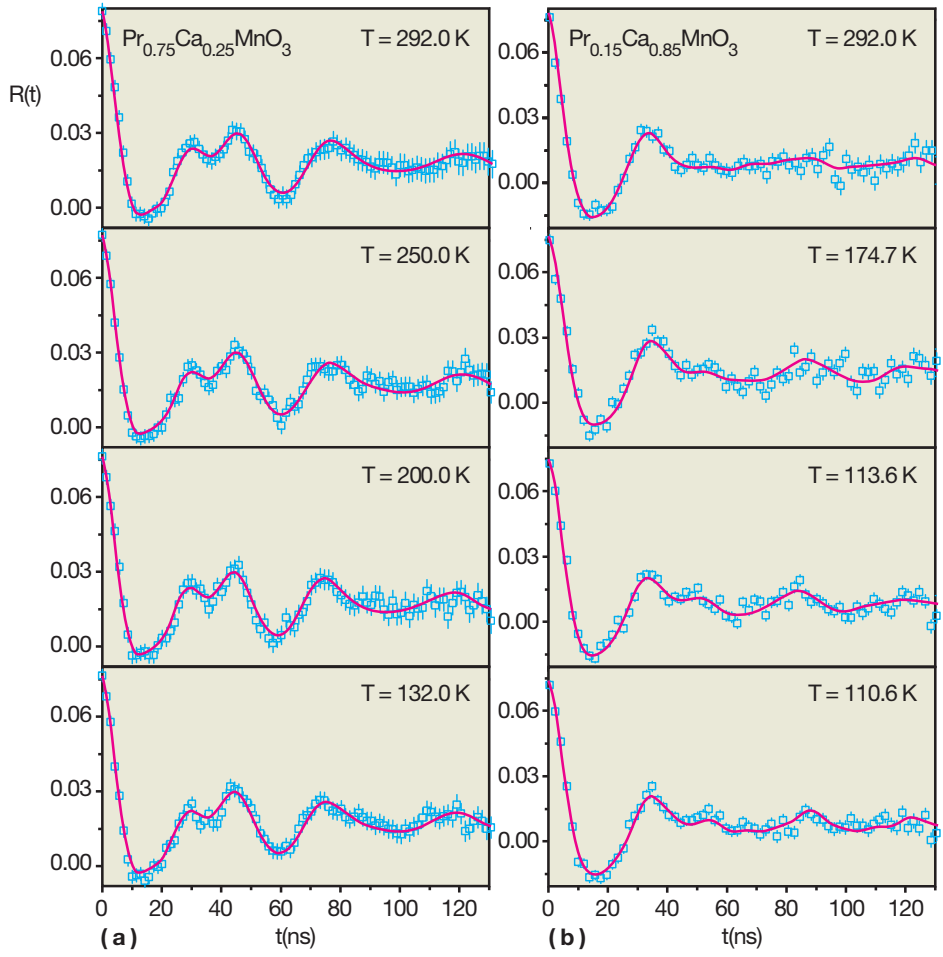

FIG. 1. Representative anisotropy functions, R(t), for different temperatures. $Pr_{0.75}Ca_{0.25}MnO_3$ sample (a) and $Pr_{0.15}Ca_{0.85}MnO_3$ sample (b).

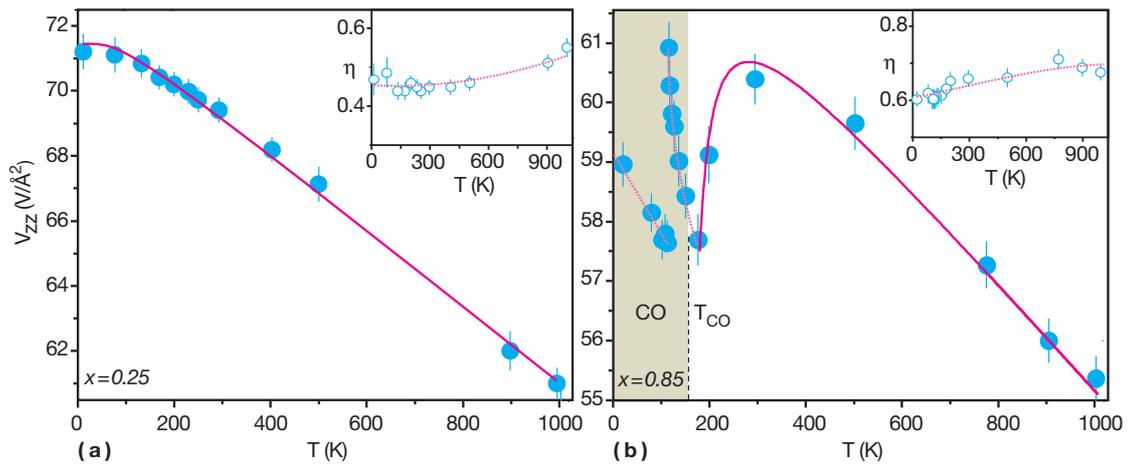

FIG. 2. Thermal dependence of the electric field gradient parameters. (a)-(b) Principal component of the electric field gradient, $V_{ZZ}$ and asymmetry parameter, $\eta$, (inset) as a function of temperature for the $Pr_{0:75}Ca_{0:25}MnO_3$ (a) and $Pr_{0:15}Ca_{0:85}MnO_3$ (b) samples. Continuous lines are fits to the data and dash lines are guides for the eye. The charge order temperature $T_{CO}$ is shown.

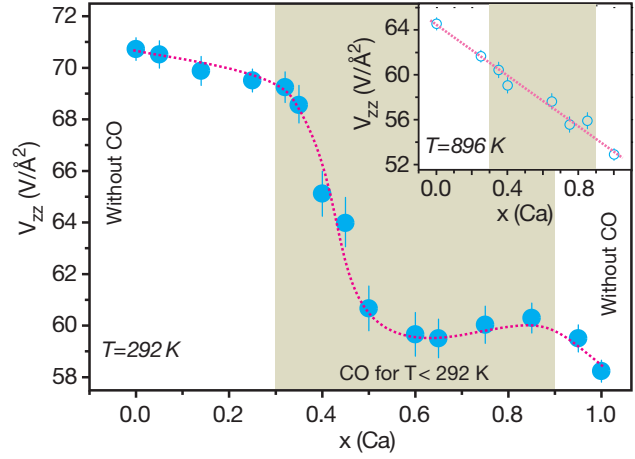

FIG. 3. Dependence of the EFG principal component ($V_{ZZ}$) on calcium content at room temperature and at 896 K (inset). Shadowed region delimits the charge order, CO, region of the phase diagram. Lines are guides for the eye.

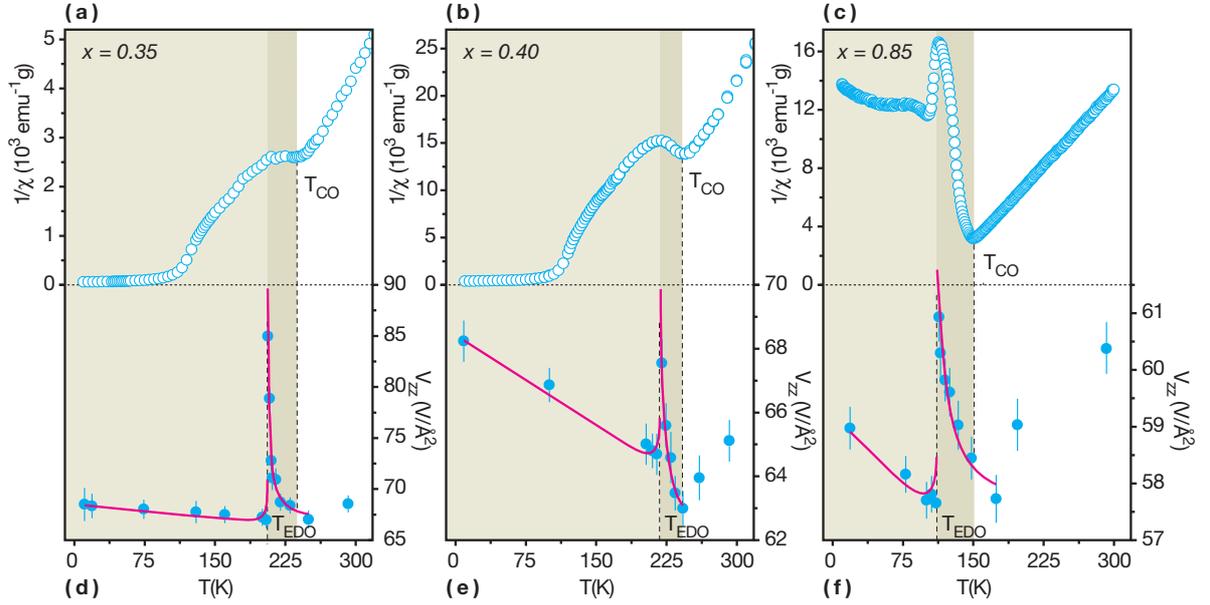

FIG. 4. Phase transitions below charge ordering. (a)-(c) Inverse magnetic susceptibility thermal dependence for $Pr_{1-x}Ca_xMnO_3$ samples with x=0.35 (a), x=0.40 (b), and x=0.85 (c). The charge order temperature, $T_{CO}$, is obtained by minimum in the inverse magnetic susceptibility. (d)-(f) $V_{ZZ}$ thermal dependence and correspondent fits below the CO transition using the Landau theory of phase transitions for x=0.35 (d), x=0.40 (e), and x=0.85 (f). $T_{EDO}$ defines the electric dipole order critical temperature obtained from the fit. The high temperature ($T>T_{CO}$) fits are not shown.